\documentstyle[12pt]{article}

\begin{document}
{\sf \begin{center} \noindent {\Large \bf An example of anti-dynamo conformal Arnold metric}\\[3mm]

by \\[0.3cm]{\sl L.C. Garcia de Andrade}\\

\vspace{0.5cm} Departamento de F\'{\i}sica
Te\'orica -- IF -- Universidade do Estado do Rio de Janeiro-UERJ\\[-3mm]
Rua S\~ao Francisco Xavier, 524\\[-3mm]
Cep 20550-003, Maracan\~a, Rio de Janeiro, RJ, Brasil\\[-3mm]
Electronic mail address: garcia@dft.if.uerj.br\\[-3mm]
\vspace{2cm} {\bf Abstract}
\end{center}
\paragraph*{}
A 3D metric conformally related to Arnold cat fast dynamo metric:
${ds_{A}}^{2}=e^{-{\lambda}z}dp^{2}+e^{{\lambda}z}dq^{2}+dz^{2}$ is
shown to present a behaviour of non-dynamos where the magnetic field
exponentially decay in time. The Riemann-Christoffel connection and
Riemann curvature tensor for the Arnold and its conformal
counterpart are computed. The curvature decay as z-coordinates
increases without bounds. Some of the Riemann curvature components
such as $R_{pzpz}$ also undergoes dissipation
 while component $R_{qzqz}$ increases without bounds. The remaining curvature
 component $R_{pqpq}$ is constant on the torus surface. The Riemann curvature invariant $K^{2}=R_{ijkl}R^{ijkl}$
 is found to be $0.155$ for the ${\lambda}=0.75$. A simple solution of Killing equations for Arnold metric yields
 a stretch Killing vector along one direction and compressed along other direction in order that the modulus of
 the Killing vector is not constant along the flow. The flow is shown to be untwisted.  The stability of the two metrics are found
 by examining the sign of their curvature tensor components. \vspace{0.5cm} \noindent {\bf PACS numbers:}
\hfill\parbox[t]{13.5cm}{02.40.Hw-Riemannian geometries}

\newpage
\section{Introduction}
 Geometrical tools have been used with success \cite{1} in
 Einstein general relativity (GR) have been also used
 in other important areas of physics, such as plasma structures in tokamaks
 as been clear in the Mikhailovskii \cite{2} book to investigate the tearing
 and other sort of instabilities in confined plasmas \cite{2}, where the Riemann metric
 tensor plays a dynamical role interacting with the magnetic field through
 the magnetohydrodynamical equations (MHD). Recently
 Garcia de Andrade \cite{3} has also made use of Riemann metric to investigate
 magnetic flux tubes in superconducting plasmas.
 Thiffault and Boozer \cite{4} following the same reasoning applied the methods of Riemann
 geometry in the context of chaotic flows and fast dynamos. In this paper we use the other tools of Riemannian geometry,
 also user in GR, such as Killing symmetries , and Ricci
 collineations , to obtain Killing symmetries in the cat dynamo metric \cite{5}. We also use the Euler equations for
 incompressible flows in Arnold metric \cite{6}. Antidynamos or non-dynamos are also important in the respect that it is important
 to recognize when a topology or geometry of a magnetic field does force the field to decay exponentially for example. As we know
 planar dynamos does not exist and Anti-dynamos theorems are important in this respect. Thus in the present paper we also obtain
 antidynamos metrics which are conformally related to the fast dynamo metric discovered by Arnold.
 Levi-Civita connections \cite{7}  are found together Riemann curvature from the MAPLE X GR tensor
 package. The paper is organized as follows: In section 2 the
 curvature and connection are found and the Euler equation is found. In section
 3 the Killing symmetries are considered. In section 4 the conformal anti-dynamo metric is presented with the new feature
 that the magnetic field decays exponentially in time along the longitudinal flux tube flow. Conclusions are presented in section 5.
 \section{Riemann dynamos and dissipative manifolds and Euler flows}
 Arnold metric can be used to compute the Levi-Civita-Christoffel connection
\begin{equation}
 {{\Gamma}^{p}}_{pz}=-\frac{\lambda}{2}\label{1}
\end{equation}
\begin{equation}
 {{\Gamma}^{q}}_{qz}=\frac{\lambda}{2}\label{2}
\end{equation}
\begin{equation}
 {{\Gamma}^{z}}_{pp}=\frac{\lambda}{2}e^{-{\lambda}z}\label{3}
\end{equation}
\begin{equation}
 {{\Gamma}^{z}}_{qq}=-\frac{\lambda}{2}e^{-{\lambda}z}\label{4}
\end{equation}
from these connection components one obtains the Riemann tensor
components
\begin{equation}
 {R}_{pqpq}=-\frac{{\lambda}^{2}}{4}\label{5}
\end{equation}
Note that since this component is negative from the Jacobi equation
\cite{7} that the flow is unstable. The other components are
\begin{equation}
 {R_{pzpz}}=-\frac{{\lambda}^{2}}{2}e^{-{\lambda}z}\label{6}
\end{equation}
\begin{equation}
 R_{zqzq}=-\frac{{\lambda}^{2}}
 {2}e^{{\lambda}z}\label{7}
\end{equation}
one may immediatly notice that at large values of z the curvature
component $(zpzp)$ is bounded and vanishes,or undergoes a
dissipative effect, while component $(zqzq)$ of the curvature
increases without bounds, component $(pqpq)$ remains constant. As in
GR or general Riemannian manifolds, to investigate singular
curvature behaviours we compute the so-called Kretschmann scalar
$K^{2}$ defined in the abstract as
\begin{equation}
 K^{2}={R}_{ijkl}R^{ijkl}=[R_{pzpz}g^{qq}g^{zz}]^{2}+ [R_{pzpz}g^{pp}g^{zz}]^{2}+[R_{qzqz}g^{qq}g^{zz}]^{2}=\frac{3}{16}
 {\lambda}^{4}
\label{8}
\end{equation}
with the value of $0.75$ for ${\lambda}$ one obtains $K^{2}=0.155$.
Which would give a almost flat on singular manifold. In GR for
example when this invariant is $\infty$ the metric is singular. This
would be a useful method to find singularities in dynamos. Let us
now compute the forced Euler equation. The forced Euler equation in
3D manifold ${\cal R}^{3}$ is
\begin{equation}
<\vec{v},\nabla>\vec{v}={\vec{F}}\label{9}
\end{equation}
where $\vec{v}$ is the speed of the flow and $\vec{F}$ is the
external force to the flow. By expressing the flow velocity in 3D
curvilinear coordinates basis $\vec{e_{i}}$ $(i,j=p,q,z)$ we obtain
\begin{equation}
(v^{i}<\vec{e_{i}},\vec{e}^{k}>{\partial}_{k})v^{l}\vec{e}_{l}=F^{k}{\vec{e}}_{k}\label{10}
\end{equation}
Since the Kr\"{o}necker delta is given by
$<\vec{e^{i}},\vec{e_{j}}>={{\delta}^{i}}_{j}$ we may write the
Euler equation in the form
\begin{equation}
(v^{k}{\partial}_{k})v^{l}\vec{e}_{l}=F^{k}{\vec{e}}_{k}\label{11}
\end{equation}
Expanding the derivative on the LHS one obtains
\begin{equation}
<\vec{v},\nabla>\vec{v}=
[v^{k}D_{k}v^{l}]\vec{e}_{l}=F^{l}{\vec{e}}_{l}\label{12}
\end{equation}
where D  is the covariant
 Riemannian derivative as defined in
reference 1. By making use of the Gauss equation
\begin{equation}
{\partial}_{k}\vec{e}_{p}={{\Gamma}^{p}}_{kl}\vec{e^{l}}\label{13}
\end{equation}
The covariant derivative can be expressed by
\begin{equation}
D_{k}v^{l}={\partial}_{k}v^{l}-{{\Gamma}^{p}}_{kl}v^{l}\label{14}
\end{equation}
Thus the Euler force equation becomes
\begin{equation}
v^{k}D_{k}v^{l}=F^{l}\label{15}
\end{equation}
Computation of the p-component of the force leads to
\begin{equation}
v^{z}{\partial}_{z}v^{p}=F^{p}\label{16}
\end{equation}
In the next section we shall compute the Killing vector equation and
yield a simple solution.
\section{Killing equations for fast dynamos}
The Killing symmetries are defined by the Killing equations
\begin{equation}
{\cal L}_{\chi}g=0\label{17}
\end{equation}
where $\vec{\chi}$ represent the Killing vector and g represents the
metric tensor. Explicitly this equation reads
\begin{equation}
[{\partial}_{l}g_{ik}]{\chi}^{l}+g_{il}{\partial}_{k}{\chi}^{l}+g_{kl}{\partial}_{i}{\chi}^{l}=0
\label{18}
\end{equation}
which explicitly reads
\begin{equation}
-{\lambda}g_{pp}{\chi}^{z}+2g_{pp}{\partial}_{p}{\chi}^{p}=0
\label{19}
\end{equation}
\begin{equation}
{\lambda}g_{qq}{\chi}^{z}+ 2g_{qq}{\partial}_{q}{\chi}^{q}=0
\label{20}\end{equation}
\begin{equation}
e^{-{\lambda}z}{\partial}_{z}{\chi}^{p}+{\partial}_{p}{\chi}^{z}=0
\label{21}
\end{equation}
\begin{equation}
e^{-{\lambda}z}{\partial}_{q}{\chi}^{p}+e^{{\lambda}z}{\partial}_{p}{\chi}^{q}=0
 \label{22}\end{equation}
Note that a very simple solution for this system can be obtained if
we put ${\chi}^{p}=c_{1}$, ${\chi}^{q}=c_{2}$, and ${\chi}^{z}=0$,
where $c_{1}$ and $c_{2}$ are constants. Since this Killing vector
has to satisfy the modulus condition
\begin{equation}
|\vec{\chi}|^{2}=g_{pp}[{\chi}^{p}]^{2}+g_{qq}[{\chi}^{q}]^{2}=[c_{1}]^{2}e^{-{\lambda}z}+[c_{2}]^{2}e^{{\lambda}z}
\label{23}\end{equation} one immediatly notices that the modulus of
the Killing vector cannot be constant along the flow, and is stretch
along the q-direction and compressed along the p-direction. In the
next section we shall analyze a new solution of MHD dynamo equation
which is conformally related to the Arnold fast dynamo metrics where
stretch and compressible behaviors of the magnetic field appear as
well.
\section{Conformal anti-dynamo metric}
Conformal metric techniques have been widely used as a powerful tool
obtain new solutions of the Einstein's field equations of GR from
known solutions. By analogy, here we are using this method to yield
new solutions of MHD anti-dynamo solutions from the well-known fast
dynamo Arnold solution. We shall demonstrate that distinct physical
features from the Arnold solution maybe obtained. The conformal
metric line element can be defined as
\begin{equation}
ds^{2}={\lambda}^{-2z}{ds_{A}}^{2}={dx_{+}}^{2}+{\lambda}^{-4z}{dx_{-}}^{2}+{\lambda}^{-2z}dz^{2}\label{24}
\end{equation}
where we have used here the Childress and Gilbert \cite{5} notation
for the Arnold metric in ${\cal R}^{3}$ which reads now
\begin{equation}
{ds_{A}}^{2}={\lambda}^{2z}{dx_{+}}^{2}+{\lambda}^{-2z}{dx_{-}}^{2}+dz^{2}\label{25}
\end{equation}
where the coordinates are defined by
\begin{equation}
\vec{x}=x_{+}\vec{e_{+}}+x_{-}\vec{e_{-}}+z\vec{e}_{z}\label{26}
\end{equation}
where a right handed orthogonal set of vectors in the metric is
given by
\begin{equation}
\vec{f}_{+}=\vec{e}_{+} \label{27}
\end{equation}
\begin{equation}
\vec{f}_{-}={\lambda}^{2z}\vec{e}_{-} \label{28}
\end{equation}
\begin{equation}
\vec{f}_{z}={\lambda}^{z}\vec{e}_{z} \label{29}
\end{equation}
A component of a vector in this basis, such as the magnetic vector
$\vec{B}$ is
\begin{equation}
\vec{B}=B_{+}\vec{f}_{+}+B_{-}\vec{f}_{-}+B_{z}\vec{f}_{z}\label{30}
\end{equation}
The vector analysis formulas in this frame are
\begin{equation}
{\nabla}=[{\partial}_{+},{\lambda}^{2z}{\partial}_{-},{\lambda}^{z}{\partial}_{z}]\label{31}
\end{equation}
\begin{equation}
{\nabla}^{2}{\phi}=[{{\partial}_{+}}^{2}{\phi},{\lambda}^{4z}{{\partial}_{-}}^{2}{\phi},{\lambda}^{2z}{{\partial}_{z}}^{2}{\phi}]\label{32}
\end{equation}
The MHD dynamo equations are
\begin{equation}
{\nabla}.\vec{B}={{\partial}_{+}}B_{+}+{\lambda}^{2z}{{\partial}_{-}}B_{-}+{\lambda}^{z}{{\partial}_{z}}B_{z}=0\label{33}
\end{equation}
\begin{equation}
{\partial}_{t}\vec{B}+(\vec{u}.{\nabla})\vec{B}-(\vec{B}.{\nabla})\vec{u}={\epsilon}{\nabla}^{2}\vec{B}\label{34}
\end{equation}
where ${\epsilon}$ is the conductivity coefficient. Since here we
are working on the limit ${\epsilon}=0$ , which is enough to
understand the physical behavior of the fast dynamo, we do not need
to worry to expand the RHS of equation (\ref{34}), and it reduces to
\begin{equation}
(\vec{u}.{\nabla})\vec{B}={\partial}_{z}[B_{+}\vec{e}_{+}+B_{-}e^{2{\mu}z}\vec{e}_{-}+B_{z}e^{{\mu}z}\vec{e}_{z}]\label{35}
\end{equation}
where we have used that
$(\vec{B}.{\nabla})\vec{u}=B_{z}{\mu}e^{{\mu}z}\vec{e}_{z}$ and that
${\mu}=log{\lambda}$. This is one of the main differences between
Arnold metric and ours since in his fast dynamo, this relation
vanishes since in Arnold metric $\vec{u}=\vec{e}_{z}$ where
$\vec{e}_{z}$ is part of a constant basis. Separating the equation
in terms of the coefficients of $\vec{e}_{+}$, $\vec{e}_{-}$ and
$\vec{e}_{z}$ respectively one obtains the following scalar
equations
\begin{equation}
{\partial}_{z}B_{+}+{\partial}_{t}B_{+}=0\label{36}
\end{equation}
\begin{equation}
{\partial}_{t}B_{-}+{\partial}_{t}B_+2{\mu}B_{-}=0\label{37}
\end{equation}
\begin{equation}
{\partial}_{t}B_{z}+{\partial}_{z}B_=0\label{38}
\end{equation}
Solutions of these equations allows us to write down an expression
for the magnetic vector field $\vec{B}$ as
\begin{equation}
\vec{B}=[{B^{0}}_{z},{\lambda}^{-(t+z)}{B^{0}}_{-},{B^{0}}_{z}](t-z,y,x+y)\label{39}
\end{equation}
From this expression we can infer that the field is carried in the
flow, stretched in the $\vec{f}_{z}$ direction and compressed in the
$\vec{f}_{-}$ direction, while in Arnold's cat fast dynamo is also
compressed along the $\vec{f}_{-}$ direction but is stretched along
$\vec{f}_{+}$ direction while here this direction is not affected.
But the main point of this solution is the fact that the solution
represents an anti-dynamo since as one can see from expression
(\ref{39}) the magnetic field fastly decays exponentially in time as
$e^{{\mu}(t+z)}$. Let us now compute the Riemann tensor components
of the new conformal metric to check for the stability of the
non-dynamo flow. To easily compute this curvature components we
shall make use of Elie Cartan \cite{8} calculus of differential
forms, which allows us to express the conformal metric as
\begin{equation}
ds^{2}={dp}^{2}+e^{4{\lambda}z}{dq}^{2}+e^{{\lambda}z}dz^{2}\label{40}
\end{equation}
or in terms of the frame basis form ${\omega}^{i}$
is
\begin{equation}
ds^{2}=({{\omega}^{p}})^{2}+({{\omega}^{q}})^{2}+({{\omega}_{z}})^{2}\label{41}
\end{equation}
where we are back to Arnold's notation for convenience. The basis
form are write as
\begin{equation}
{\omega}^{p}=dp \label{42}
\end{equation}
\begin{equation}
{\omega}^{q}=e^{{\lambda}z}dq \label{43}
\end{equation}
and
\begin{equation}
{\omega}^{z}=e^{{\frac{\lambda}{2}}z}dq \label{44}
\end{equation}
By applying the exterior differentiation in this basis form one
obtains
\begin{equation}
d{\omega}^{p}=0 \label{45}
\end{equation}
\begin{equation}
d{\omega}^{z}=0 \label{46}
\end{equation}
and
\begin{equation}
d{\omega}^{q}={\lambda}e^{-{\frac{\lambda}{2}}z}{\omega}^{z}{\wedge}{\omega}^{q}
\label{47}
\end{equation}
Substitution of these expressions into the first Cartan structure
equations one obtains
\begin{equation}
T^{p}=0={{\omega}^{p}}_{q}{\wedge}{\omega}^{q}+
{{\omega}^{p}}_{z}{\wedge}{\omega}^{z}\label{48}
\end{equation}
\begin{equation}
T^{q}=0={\lambda}e^{-{\frac{\lambda}{2}}z}{\omega}^{z}{\wedge}{\omega}^{q}+{{\omega}^{q}}_{p}{\wedge}{\omega}^{p}+{{\omega}^{q}}_{z}{\wedge}{\omega}^{z}
\label{49}
\end{equation}
and
\begin{equation}
T^{z}=0={{\omega}^{z}}_{p}{\wedge}{\omega}^{p}+{{\omega}^{z}}_{q}{\wedge}{\omega}^{q}
\label{50}
\end{equation}
where $T^{i}$ are the Cartan torsion 2-form which vanishes
identically on a Riemannian manifold. From these expressions one is
able to compute the connection forms which yields
\begin{equation}
{{\omega}^{p}}_{q}=-{\alpha}{\omega}^{p}\label{51}
\end{equation}
\begin{equation}
{{\omega}^{q}}_{z}={\lambda}e^{-{\frac{\lambda}{2}}z}{\omega}^{q}
\label{52}
\end{equation}
and
\begin{equation}
{{\omega}^{z}}_{p}={\beta}{\omega}^{p} \label{53}
\end{equation}
where ${\alpha}$ and ${\beta}$ are constants. Substitution of these
connection form into the second Cartan equation
\begin{equation}
{R^{i}}_{j}={R^{i}}_{jkl}{\omega}^{k}{\wedge}{\omega}^{l}=d{{\omega}^{i}}_{j}+{{\omega}^{i}}_{l}{\wedge}{{\omega}^{l}}_{j}
\label{54}
\end{equation}
where ${R^{i}}_{j}$ is the Riemann curvature 2-form. After some
algebra we obtain the following components of Riemann curvature for
the conformal antidynamo
\begin{equation}
{R^{p}}_{qpq}= {\lambda}e^{-{\frac{\lambda}{2}}z}\label{55}
\end{equation}
\begin{equation}
{R^{q}}_{zqz}= \frac{1}{2}{\lambda}^{2}e^{-{\lambda}z}\label{56}
\end{equation}
and finally
\begin{equation}
{R^{p}}_{zpq}= -{\alpha}{\lambda}e^{-{\frac{\lambda}{2}}z}\label{57}
\end{equation}
We note that only component to which we can say is positive is
${R^{p}}_{zqz}$ which turns the flow stable in this q-z surface.
This component also dissipates away when $z$ increases without
bounds, the same happens with the other curvature components
\cite{8}.

\section{Conclusions}
 In conclusion, we have used a well-known technique to find solutions of Einstein's field equations of gravity
 namely the conformal related spacetime metrics to find a new anti-dynamo solution in MHD nonplanar flows.
 The stability of the flow is also analysed by using other tools from
 GR, namely that of Killing symmetries. Examination of the  Riemann curvature components enable one to
 analyse the stretch and compression of the dynamo flow. The Killing symmetries can be used in near future to classify the
 dynamo metrics in the same way they were useful in classifying general relativistic solutions of Einstein's gravitational
 equations in four-dimensional spacetime \cite{1}. \section*{Acknowledgements}
 Thanks are due to CNPq and UERJ for financial supports.

\end{document}